\documentclass[10pt,english,sigconf,authoryear,nonacm]{acmart}
\usepackage[T1]{fontenc}
\usepackage[latin9]{inputenc}
\usepackage{array}
\usepackage{verbatim}
\usepackage{float}
\usepackage{amsmath, amsthm}
\usepackage{graphicx}

\makeatletter

\providecommand{\tabularnewline}{\\}
\floatstyle{ruled}
\newfloat{algorithm}{tbp}{loa}
\providecommand{\algorithmname}{Algorithm}
\floatname{algorithm}{\protect\algorithmname}

\newenvironment{lyxlist}[1]
	{\begin{list}{}
		{\settowidth{\labelwidth}{#1}
		 \setlength{\leftmargin}{\labelwidth}
		 \addtolength{\leftmargin}{\labelsep}
		 }}
	{\end{list}}
\usepackage{algolyx}

\usepackage{url}
\usepackage{booktabs}
\usepackage{tikz}
\usetikzlibrary{positioning,arrows,calc,fit,automata}

\usepackage{pgfplots}
\DeclareUnicodeCharacter{2212}{}
\usepgfplotslibrary{groupplots,dateplot}
\usetikzlibrary{patterns,shapes.arrows}
\pgfplotsset{compat=newest}

\usepackage{nccmath}
\usepackage{mathtools}
\usepackage[skins]{tcolorbox}

\usepackage{multicol}

\AtBeginDocument{
  
}

\makeatother

\usepackage{babel}
\begin{document}
\title{Towards Detecting Inauthentic Coordination in Twitter Likes Data}

\author{Laura Jahn and Rasmus K. Rendsvig}
\email{}

\affiliation{%
  \institution{Center for Information and Bubble Studies, Department of Communication, University of Copenhagen}
  \streetaddress{}
  \city{}
  \state{}
  \country{}
  \postcode{}
}

\date{}
\begin{abstract}
\noindent Social media feeds typically favor posts according to
user engagement. The most ubiquitous type of engagement (and the type
we study) is \emph{likes}. Users customarily take engagement metrics
such as likes as a neutral proxy for quality and authority. This incentivizes
like manipulation to influence public opinion through \emph{coordinated
inauthentic behavior }(CIB). CIB targeted at likes is largely unstudied
as collecting suitable data about users' liking behavior is non-trivial.
This paper contributes a scripted algorithm to collect suitable liking
data from Twitter and a collected 30 day dataset of liking data from
the Danish political Twittersphere \texttt{\#dkpol}, over which we
analyze the script's performance. Using only the binary matrix of
users and the tweets they liked, we identify large clusters of perfectly
correlated users, and discuss our findings in relation to CIB.
\end{abstract}


\maketitle

\section{Introduction}

Algorithmically curated social media feeds favor posts according to
user engagement. The most ubiquitous type of engagement (and the type
we study) is \emph{likes} \cite{Torres-Lugo_Likes_Manip_deletions}.
A post\textemdash a tweet, a shared news article, a video, a meme,
etc.\textemdash may be highlighted e.g. by being placed highly on
users' news feeds. Users customarily take engagement metrics such
as likes as a neutral proxy for quality and authority \cite{Torres-Lugo_Likes_Manip_deletions,MCSpaper}.
This incentivizes\emph{ influence operations} to misrepresent, mislead
or manipulate opinion dynamics online \cite{nizzoli2021coordinated}.
Such media manipulation tactics have been labeled \emph{coordinated
inauthentic behavior }(CIB) \cite{Pacheco2021Coordinated,Orabi2020,Goerzen2019,Ferrara2017,Ferrara:2016,Giglietto2020coordinated}.
Influence operations and CIB may thus shape public opinion and political
discourse through \emph{attention hacking}, the act of exploiting
platforms' content sorting algorithms to highlight certain information
items to users. This highlights the societal need to address CIB-caused
misrepresentation of  political views and the spread of harmful low-quality
content and misinformation in the online public sphere \cite{MCSpaper}.

To effectively push narratives on social media, influence operations
resort to \emph{coordinated} groups of accounts rather than individual
accounts \cite{Kirn2022,Khaund2022_Socialbotcoord}. This has, for
example, led to the establishment of a marketplace for vendor-purchased
engagement \cite{Torres-Lugo_Likes_Manip_deletions,ikram2017_fblikefarms}
and metric inflation through coordinated social bots. The behavior
dictated by an influence operation is labeled \emph{inauthentic} as
it may not reflect the personal beliefs of the instructed user accounts,
as these accounts may be run by algorithmic amplifiers such as automated
bots or humans according to a supplied protocol \cite{duan2022botsIndiana}.

CIB targeted at one-click reactions such as likes is largely unstudied
as collecting data about users' liking behavior around a specific
political discourse is non-trivial due to the lack of access to platform
data for researchers or severe API rate restrictions that prevent
collecting comprehensive datasets. The first main contribution of
this paper is a script to collect comprehensive data on liking users
from Twitter\emph{.} The second main contribution is a dataset collected
with the script. The dataset contains a month-long survey of liking
user behavior from the Danish political Twittersphere, collected through
the hashtag \texttt{\#dkpol} (``DenmarK POLitics''). Under this
hashtag, citizens, organizations, politicians and journalists from
across the political spectrum air, discuss and orientate themselves
about current debates in Danish politics. It is \emph{the} centralized,
place-to-be source of information on the debates of the day. The hashtag
thus seems a likely candidate for inauthentic coordination, if one
seeks to increase the Danish public sphere's attention on some topic.
We use the dataset first to evaluate the effectiveness of the script,
and second as basis for a case study of liking users behavior with
the aim to determine if the simple liking data has sufficient structure
to serve as an entry point for the detection of CIB. We argue that
it does.

Using a running survey approach, the script retrieves IDs of the most
recent liking users of tweets satisfy a specified text query (e.g.
a keyword or hashtag of a chosen political debate), timing retrievals
by taking into account Twitter set rate limits of the public v2 API
for Academic Research Access. The script can retrieve far more comprehensive
sets of liking user IDs than are available through the default public
and commercial tools of the Twitter APIs and Decahose stream. To the
best of our knowledge, the resulting data is the first to contain
comprehensive collections of user-IDs of liking users. The dataset
thus advances the specialized field of studying one-click reaction-based
CIB.

The script's point of departure for data collection is the survey
of an online\emph{ discourse} around a \emph{domain} (e.g. a hashtag)
instead of a survey of a preselected group of users. Hence, data collection
does not require any prior knowledge about potentially coordinated
users nor does subsequent data analysis necessarily require the retrieval
of additional account data. When identifying coordination of likes
given such concise data, one immediately grasps firstly which specific
tweet(s) a potential influence operation is targeted at, and secondly
which users are involved in the metric inflation (this is in contrast
to existing methods for collecting retweeting user IDs, cf. Sec.~\ref{subsec:Related-work}
below). If desired, additional account information may then be rehydrated
via public APIs. The focus of the collected data and following applications
is thus rather on identifying the \emph{effects} of CIB inflating
specific tweets. These effects may be more robust to changes in the
evolution of algorithmic amplifiers, social bots and cyborgs, that
with varying degrees of automation increasingly emulate authentic
users. Our data and applications are not dependent on individual account
features nor time-synchronous actions but only on the like behavior
towards an observed tweet.

We analyse the dataset in a case study of \texttt{\#dkpol}, mainly
to illustrate that the liking behavior data has sufficient structure
to serve as a point of entry for detection of CIB. Pre-processing
the data points into a simple binary and sparse tweet/like matrix
suffices to detect like-coordinated accounts without relying on textual,
temporal, nor training data (see Sec.~\ref{subsec:CaseStudy:}),
a topic that has previously gone unstudied. We undertake two simple
analyses: First, we group users by the toughest clustering criteria
of complete equality of their like profiles. Under this very strict
criteria, we identify several large perfectly correlated groups, including
likes we purchased from online vendors. Notably, we detect the vendor-purchased
CIB and more perfectly correlated groups of users despite the users
not being particularly active (one like suffices), so without any
requirement that they have liked aggressively. Second, we show that
these groups can be visualized using the first two dimensions in a
dimensionality-reduced space using the first two eigenvectors of a
Singular Value Decomposition of the tweet/like matrix.

Given a lack of ground truth, we cannot be sure the perfectly coordinated
clusters we detect (other than the vendor-purchased groups) are artifacts
of CIB. We do believe that the natural correlation is unlikely enough
that the groupings raise red flags, warranting further inspection,
out of scope of this case study. Our methods may thus serve as pre-studies
for bot detection and the application of fact checkers \cite{Kirn2022}.

We make our resources available to the research community, including
the raw datapoints complemented with timestamp data (tweet text must
be rehydrated per Twitter data sharing policies) and pre-processed
user-like data matrices, the scripts used for data collection, for
data pre-processing, for evaluation of the completeness of a collected
dataset, and for clustering and visualization. Data and scripts are
available on Harvard Dataverse \cite{Dataverse} and the data collection
script is additionally available at the public GitHub repository \emph{Get-Twitter-Likers-Data}
\cite{JahnRendsvig22Git_likers}.

\subsection{Related Work\label{subsec:Related-work}}

Social media users have a plethora of available action types \cite{Magelinski2022},
many of which may be used in coordinated fashions. E.g., users may
coordinate using a specific hashtag, posting a specific URL, tweet,
image or mention, or coordinate replies, shares or reactions to existing
content. As coordination is not visible when inspecting accounts in
isolation, research on CIB has turned to study the collective behavior
of groups, with similarities between users serving as a proxy for
coordination. Studies have analysed similarities between users posting
similar \emph{content} \cite{Ahmed2013,AlKhateebAgrawal2016_coord,Pacheco2021Coordinated,Kirn2022},
users having similar\emph{ friends and followers} \cite{nizzoli2021coordinated},
and having similarly \emph{timed activities }(e.g.,\emph{ }\cite{Giglietto2020coordinated,Giglietto2020coordinated2,weber2021amplifyingcoord,Magelinski2022,Cresci2018,Cresci2016,giatsoglou2015nd}).
Few studies have looked directly at coordination in one-click reactions
such as liking.

Liking is a one-click engagement where users may select one option
from a short pre-defined list as their `reaction' to a post, with
users' choices typically summed and presented as a quantified metric
beneath the item. Reactions include perhaps most famously Facebook's
original `Like', the hearts/likes on Instagram, TikTok and Twitter,
and Reddit's up- and downvotes. Sharing and retweeting may also be
taken as a one-click reaction on any of these platforms.

Importantly, these reactions inform the platforms' algorithmic content
sorting, thus steering users' attention. With attention metrics such
as likes being widely used as a proxy for quality and authority, manipulating
like counts becomes incentivized for the sake of increased exposure,
influence, and financial gain \cite{Torres-Lugo_Likes_Manip_deletions}.
High engagement counts may be perceived as a trust signal about the
content \cite{Metaxas2015} and as a positive crowd reaction aiding
content to broadcast and to trend \cite{dutta2018retweet}. Once
trending, high engagement counts in likes and shares make users more
likely to engage with low-credibility content instead of fact-checking
questionable posts \cite{avram2020exposure}. Scholars have stressed
that to fight disinformation campaigns, it is less effective to look
at the pushed content (e.g., hashtags, URLs, memes, etc.) and more
effective to look at the coordinated content pushing \emph{behaviors}
\cite{Kirn2022}.

\paragraph{Related work on coordinated retweeting}

To push stories online, retweeting and inflation of the retweet metric
attracts manipulation. Several recent papers look at retweeting as
a coordination dimension.

Dutta et al. \cite{dutta2018retweet} investigate non-synchronized,
collusive retweeters ($n<1,500$) involved with \emph{blackmarket}
services. Such collusive retweeters re-share the tweets of other blackmarket
customers to earn credits. The authors use a human annotated dataset
and supervised machine learning methods leveraging features such as,
e.g., user activity or social network characteristics to distinguish
between \emph{customers} and \emph{genuine retweeters}, later extended
to detect \emph{paying customers} \cite{Dutta2020}. Building on these
works, Arora et al. \cite{arora2020analyzing} analyze user representations
to improve the performance of detecting blackmarket customers while
Chetan et al. \cite{chetan2019corerank} develop an unsupervised approach
to detect collusive blackmarket retweeters leveraging, for example,
the merit of tweets and timing of retweets analyzed through a bipartite
tweet-user graph.

Schoch et al. \cite{Schoch2022} study time-synchronous co-retweeting
(and co-tweeting) as a trace of coordination to detect astroturfing
campaigns given a dataset released by Twitter consisting of tweets
by accounts that Twitter classified as being involved in hidden information
campaigns. The authors filtered the data and only looked at campaigns
with more than $50,000$ tweets and users that tweeted at least $10$
times in the observation period. They do so by analyzing timing and
centralization of coordination. The approach rests on the assumption
that it seems implausible that repeated co-retweeting and co-tweeting
happens without centralized coordination (e.g., one actor controlling
multiple accounts) in a small time window of $1$ minute up to 8 hours.
Increasing the temporal window beyond that yields higher false positive
rates in flagging astroturfing accounts. The study builds a co-(re)tweeting
graph by drawing an edge between two users that (re)tweet the same
post within a minute, but only if this can be observed more than 10
times. While the authors rightfully claim that co-retweeters and co-tweeters
can be rehydrated from a Twitter dataset, it remains a necessity that
one has selected a list of users prior to dataset construction. Some
knowledge over the presence of astroturfers is hence necessary a priori:
Their approach presupposes to have a list of (suspicious) users instead
of embarking on detection given an observable effect.

Similarly concerned with co-retweeting, Graham et al. \cite{Graham2020}
searches for evidence of bots in $>25$ million retweets of $>2.5$
million tweets, collected over the course of 10 days, containing COVID-related
hashtags. The authors create a user-user `\emph{bot-like' co-retweet
network} of $>5,000$ Twitter accounts that \emph{frequently} co-retweeet
the same tweets within a time window as small as $1$ second, followed
by manual inspection of the connected components.

Pacheco et al. \cite{Pacheco2021Coordinated} take a high number
of overlapping retweets (co-retweeting) as a coordination trace and
construct a bipartite network between retweeting accounts and retweeted
messages, filtering for accounts that logged at least 10 retweets.
The authors represent users with TD-IDF weighted vectors containing
the retweeted IDs. The weighting discounts the contributions of popular
tweets. The projected co-retweet graph is then established via the
cosine similarity between the account vectors. Using a hard threshold,
they only keep the most suspicious 0.5\% edges leaving them with a
coordinated set of users. The analysis is conducted on an anonymized
dataset from DARPA SocialSim containing identified Russian disinformation
campaigns, collected from Twitter using English and Arabic keywords.
Messages that were identified as coordinated are no longer publicly
available.

Interested in how well network communities hide from coordination
detection, Weber et al. \cite{weber2021amplifyingcoord} study retweets
using a latent coordination network. When members of a group retweet
each others' posts, detection of the involved accounts becomes easy,
as the accounts are connected via an edge. The larger the detected
coordinated community, the greater the likelihood that members would
retweet other members. Notably, the authors find that large groupings
of accounts in the Twitter curated dataset, believed to be involved
in influence operations, hide well with low internal retweet ratios,
and that also official political accounts seem to refrain from being
involved in self-retweeting.

Adopting the network approach \cite{nizzoli2021coordinated,Pacheco2021Coordinated,weber2021amplifyingcoord},
Tardelli et al. \cite{TardelliNizzoliCresci_23_tempCoord} model \emph{evolving}
coordinated retweet communities. This work explores that users may
belong to different coordinated groups at different points in time.
Using the Jaccard similarity measure, the authors compare influx and
outflux into and out of communities at each time step. The resulting
temporal networks and dynamic community detection identifies many
coordinated communities and highlights the relevance of temporal nuances
of coordination.

Instead of leveraging graph-based techniques, Mazza et al. \cite{mazza2019rtbust}
only require the timestamps of retweets and the retweeted tweets for
each account, and not, e.g. full user timelines. Their work investigates
temporal and synchronous retweeting patterns. The collected data spans
short of $10$ million Italian retweets from $>1.4$ million distinct
users collected over the course of two weeks. The collected data is
filtered for human-like retweet activity between $2$ and $50$ times
per day and excludes fully automated, benign retweet bots with high
retweeting activity, resulting in a dataset with $63,762$ distinct
users. Manual annotation of a subset of the data ($1,000$ users)
serves as a ground truth. Given a user and their retweet history,
the authors first visualize different temporal retweet patterns by
plotting the timestamp of the original tweet against the timestamp
of the retweet in a scatterplot. With a granularity of seconds, the
authors compress timestamp data into per user time series vectors
containing time information if the user retweeted a given tweet at
a given time, and 0 otherwise. The resulting series remains sparse
as users usually only retweet once every few minutes. To reduce sparsity,
the data is then compressed employing a sequence compression scheme.
Using automatic unsupervised feature extraction, the work exploits
that synchronous and coordinated users will be grouped densely together
in the feature space, in contrast to heterogeneous human behavior.
The authors apply dimensionality reduction techniques and deep neural
networks and eventually hierarchical and density-based clustering.
Users that are clustered and not treated as noise (i.e., not clustered)
are labeled as bots. Users clustered together are then thought of
as bots acting in a coordinated and synchronous fashion.

\paragraph{Related work on coordinated liking.}

Despite likes being a commonly adopted and an easily manipulatable
mechanism, research on CIB more narrowly targeted at likes is quite
scarce:

Border-lining relevancy are studies on purchased likes not of posts,
but of \emph{pages} and \emph{followers} on Facebook and Twitter \cite{ikram2017_fblikefarms,Cristofaro2014_fblikes,beutel2013copycatch,aggarwal2015Twitterfakefol}.
Studying page like or follower farms \cite{Cristofaro2014_fblikes,aggarwal2015Twitterfakefol},
these works develop supervised classifiers using demographic, explicitly
temporal, and social characteristics \cite{ikram2017_fblikefarms,aggarwal2015Twitterfakefol}.
Notably, Ikram et al. \cite{ikram2017_fblikefarms} find their bot
classifier has difficulty detecting like farms that mimick regular
like-spreading over longer timespans, i.e. deliver likes slowly, without
high temporal synchronization, and with lower like counts per account.

Beutel et al. \cite{beutel2013copycatch} study coordinated and
time-synchronized attempts to inflate likes on Facebook pages. Their
unsupervised method, developed with data from inside Facebook, detects
ill-gotten likes from groups of users that coordinate to like the
same page around the same time, leveraging temporal data explicitly.
The authors follow a graph-based approach, draw a bipartite graph
between users and pages noting down the time at which each edge was
created. They then apply co-clustering looking for users liking the
same pages at around the same time. Since \cite{beutel2013copycatch}'s
approach depends on timing and is designed to detect synchronous likes
in a ``single burst of time'', \cite{ikram2017_fblikefarms} find
that \cite{beutel2013copycatch}'s approach, too, suffers large false
positive errors in detecting liking accounts that mimick regular users
and deliver likes more slowly.

While the Facebook like button is the same whether it regards a page
or a post, page likes inflation differs in the mechanism from post
like inflation. Liking a page on Facebook entails ``following''
the account, subscribing to new account posts. Thus, this kind of
coordinated metric inflation may not catapult a single \emph{post}
to the top of an algorithmically curated newsfeed but creates the
illusion of a popular \emph{account}.

Directly about reactions to posts is Torres-Lugo et al.'s \cite{Torres-Lugo_Likes_Manip_deletions}
study of metric inflation through strategic deletions on Twitter.
They analyze coordination in repetitive \emph{(un)liking} on \emph{deleted}
tweets in influence operations that seek to bypass daily anti-flooding
tweeting limits. From a collection point of view, looking at unlikes
is a smart move, as this data is in fact available to purchase from
Twitter. Alas, the approach is inapplicable to tweets that remain
online, such as those central to CIB-based influence operations that
push narratives through political astroturfing \cite{Schoch2022}.

Also in the related field of bot detection has the detection of bots
designed to engage through reactions gone unstudied, perhaps due to
data restrictions. For a systematic review of the bot detection literature,
see \cite{Orabi2020}.

\subsection{Empirical Problems}

Group-based detection methods are promising ``in the arms race against
the novel social spambot'' \cite{cresci2017paradigm}. Yet empirical
research meets challenges in this domain. The following three problems
highlight the need for a feasible data collection script and findable
datasets for researchers to develop and test methods to address CIB
targeted at reactions online.

\paragraph{Time-sensitivity.}

First, empirical social media studies of coordinated online accounts
remain problematic to replicate and reproduce due to time-sensitivity
of the relevant data \cite{Martini2021}. Attempts to collect the
same data twice are likely to fail, as traces of coordination may
be altered or deleted after an influence operation was concluded.
While e.g. Twitter grants generous academic research access to historic
tweets through their API, accounts involved in CIB may evade detection
e.g. by changing handle, so they are no longer retrievable in their
original appearance \cite{Torres-Lugo_Likes_Manip_deletions}. The
shortcomings in data reproducibility make CIB/bot detection frameworks
difficult to compare, as these typically require live data access
\cite{Martini2021}.

\paragraph{Data availability.}

Second, data availability limits research \cite{Martini2021,Disinfresearch-agenda-2020,misinfo_data,Giglietto2020coordinated}.
Large scale studies may simply be impossible due to data access restrictions
\cite{Martini2021,Disinfresearch-agenda-2020,misinfo_data}. Specifically
data concerning users' reactions is very difficult for researchers
to obtain: none of the currently existing datasets include it,\footnote{See e.g. Indiana University's Bot Repository, a resourceful, centralized
repository of annotated datasets of Twitter social bots \cite{Botometer}.} Twitter's transparency reports do not include information of liking
or retweeting users \cite{Twittertransp}, and neither Meta, Twitter
nor Reddit supply this data in necessary scope \cite{Disinfresearch-agenda-2020,misinfo_data}.

Among the platforms with APIs for academic purposes, only Twitter
releases user-IDs of (public) profiles that have liked or retweeted
a given tweet. Twitter does not give direct access to \emph{comprehensive}
lists of such IDs, but only releases the user-IDs of the 100 \emph{most
recent} liking/retweeting users of any single post. Further restrictive,
at most 75 such lists may be requested per 15 minutes. For some Twitter
environments, these restrictions may be balanced by using a suitably
timed algorithm, cf. below. For huge political hashtags like \texttt{\#MakeAmericaGreatAgain}
or \texttt{\#Brexit} where CIB-based influence operations may most
be feared to be in play, current data restrictions make it practically
impossible to obtain a complete picture of liking and retweeting behavior.
Twitter's commercial Decahose API stream lists 100\% of liking user-IDs,
but only of a \emph{random} 10\% sample of all tweets, making a targeted
analysis of a specific political discourse impossible \cite{Decahose}.

\paragraph{Ground truth.}

Third, there is an issue with lacking ground truth as researchers
have no access to the empirical truth about accounts engaged in coordinated
inauthentic behavior. Qualified guesses can be made based on suspicious
similarities in behavior or profile features, but \emph{de facto},
it remains unknown whether two users' actions are authentically correlated
or inauthentically coordinated, or how many (partially) automated
accounts exist in a total population \cite{Magelinski2022,Martini2021,beutel2013copycatch,Chavoshi2017_unsupervised}.

Specifically for reaction-based CIB, it seems infeasible to create
a labeled dataset that even \emph{approximates} the ground truth:
labeling accounts individually e.g. via crowd-sourcing or the well-established
bot classifier \emph{Botometer} will likely fail as single accounts
will often seem inconspicuous \cite{mazza2019rtbust}.\emph{ Botometer}'s
feature-based approach considers accounts one at a time and does therefore
not pick up on group anomalies based on suspicious similarity \cite{Yang2019,Yang2020}.
Especially when it comes to coordinated liking behavior, Botometer's
feature ``favourites\_count'' (the number of likes a user has delivered)
predicts less bot-like behavior, the higher the count is \cite{Yang2020},
thus undermining the attempt to identify coordinated liking. For purposes
of studying liking behavior in concert at a collective level \cite{Magelinski2022,Grimme2018CoordUnsupervOwnBots,Yang2019,Yang2020},
data availability restrictions make collective labeling impossible.

Instead of relying on (an approximation of) a ground truth, groups
of users may be labeled as suspicious, e.g. in terms of graph structure
\cite{Magelinski2022,beutel2013copycatch}, contextually validated
via manual inspection and individual confession by the original poster
\cite{Grimme2018CoordUnsupervOwnBots,mazza2019rtbust}, through NLP
of the content promoted \cite{nizzoli2021coordinated,Chavoshi2017_unsupervised},
or compared to behavior of experimental vendor-purchased metric inflation
\cite{ikram2017_fblikefarms,aggarwal2015Twitterfakefol}, as we do
in the case study in Sec.~\ref{sec:Case-Study:-Data}.

\section{Data Collection}

To collect a comprehensive dataset needed to identify coordinated
inauthentic liking behavior, we scripted an algorithm that makes effective
use of the data limits set by Twitter. Here, we aim to give an intuition
of the implementation and workings of the data collection algorithm.
We then present its pseudocode.

\subsection{Data Collection Script: Intuition}

In short, the script surveys Twitter for tweets falling under a \emph{textual
query }during a live \emph{observation period} (e.g. $30$ days).
During the observation period, with a fixed time interval $p$ (e.g.
every 5 min.), the script executes a \emph{pull}. Each pull loop contains
four steps:
\begin{enumerate}
\item It logs tweets posted since the last pull that satisfy the query,
and their current number of likes (\emph{like count}).
\item It updates the logged like count of previously logged tweets. Only
tweets that are recent enough are tracked in this way (e.g., posted
within the last $48$ hours).
\item For each logged tweet, it compares the tweet's new like count to its
like count \emph{at the last pull where its liking users were requested
}($0$ if the liking users have never been requested)\emph{.} Call
the numerical difference between these two like counts the tweet's
\emph{delta}.
\item It requests the $100$ most recent liking users of the top $n$ tweets
with the highest delta above a set threshold (e.g., has minimum $25$
new likes).
\end{enumerate}
At the end of the observation period and once every logged tweet is
no longer tracked, the liking users of all logged tweets is requested
a final time (in timed batches). The script also allows pulling retweeting
users in the pull loop. The logic is the same. Pulling liking and
retweeting users draws on separate pools of request resources.

To raise the chance of a complete data set\textemdash one that has
not missed any liking users\textemdash it is preferable to set the
tweet track time as long as possible, the pull interval $p$ as short
as possible, and the number of top $n$ tweets checked to its maximum.
Alas, this will often lead to request request shortage.

Twitter's request limits entail that the parameters of the script
have to be balanced carefully. For example, a query with 10.000 new
tweets a day, each tweet tracked for 24 hours at 5 minute intervals
uses 8.640.000 tweet-requests over a 30 day period. Twitter allows
10.000.000. The same parameters but a query with 12.000 tweets/24h
uses 10.3680.000 tweet-requests. Hence, the pull interval and the
track time must balanced with respect to the query volume. Additionally,
the pull interval $(p)$ and the number of requests used per pull
$(n)$ must also be balanced with respect to the liking frequency
\emph{and }the activity under the query. Given the 75 likers-requests
available per 15 minutes, there are two extremes (if one plays it
safe; see further below): a short pull interval of $p=\frac{15\text{ min.}}{75}=12$
seconds, each pull getting the likers the top $n=1$ tweet and a long
pull interval of $p=15$ min., each pull getting the likers of the
top $n=75$ tweets. The former lowers the risk of missing out on likers
during rapid hours, but burns through many more tweet-requests per
hour, counting against the 10.000.000 limit. Long pull intervals,
on the other hand, raise the risk of missing put on liking users.%

The script allows extending the Twitter request resources by the inclusion
of multiple bearer tokens. If working in a team where multiple members
have Academic Research access to Twitter, all their bearer tokens
may be included. The script then cycles through them, using one per
pull loop.

Finally, the pull loop is written in Python 3, and is run through
a shell script that resumes it from the point of failure in case of
Twitter connection errors, e.g. caused by an over-use of requests
or a network disruptions. This means the script allows \emph{not }playing
safe with request resources, most notably with the pull interval $p$
and the number of likers-requests used per pull, $n$. Playing it
unsafe allows for some flexibility. One may e.g. set $p=3$ min. and
$n=30$ if one trusts that the actual distribution of tweets and likes
is unlikely to break the request limit but wants to readily sacrifice
more than the safe amount of requests in case of an activity surge.

\subsection{Script: Details and Pseudocode\label{subsec:algorithm}}

The algorithm is parameterized by three time periods. First, $observationtime$
is the length of data collection (e.g. 24 hours, or 1 month), without
restriction: with properly set parameters, one can span 1 month, after
which request limits reset, making it extendable. The $observationtime$
starts at a point in time ($startpoint$). Second, $pullinterval$
defines a sleep period between the conclusion of one pull and the
initiation of the next. The shorter it is, the finer the temporal
resolution and the lower the risk of missing any liking users, but
also the higher the request usage. Third, $tracktime$ specifies how
long a tweet is monitored for new likes and retweets after it is posted
(e.g., each tweet is tracked for 1 hour, or $48$ hours). To collect
full data for all tweets posted in $observationtime$, the total scraping
time amounts to $observationtime+tracktime$.

The algorithm is split into two steps, Alg.~\ref{alg:Part-1-perpetualcollection}
and Alg.~\ref{alg:Part-2-=00005Bfinalharvest=00005D}, with Alg.~\ref{alg:Part-1-perpetualcollection}
undertaking most of the work, and collects data from Twitter using
the Academic Research access API (ARA). ARA provides significant data
scraping resources to researchers that are, however, subject to rate
limits and request caps specified by Twitter in an advance to manage
server requests. Among others, but most notably, requesting liking
users from ARA always returns the most recent $100$ liking users
of a given tweet in question. Furthermore, this request can only be
made $req.rate.lim=75$ times per $15$ minutes. As tweets routinely
get more that $100$ likes in total, a dataset that contains an as
complete as possible set of identifiable liking users must live-log
liking users runningly.

This is accomplished in Alg.~\ref{alg:Part-1-perpetualcollection},
which runs from $startpoint$ to $endpoint:=startpoint+observationtime+tracktime$.
At $endpoint$, Alg.~\ref{alg:Part-2-=00005Bfinalharvest=00005D}
runs. It completes a final harvest of liking users by requesting the
100 most recent liking users from all logged tweets. This is especially
relevant for those tweets with low like counts de-prioritized in Alg.~\ref{alg:Part-1-perpetualcollection}.

Between $startpoint$ and $endpoint$, Alg.~\ref{alg:Part-1-perpetualcollection}
performs a \emph{pull }every $pullinterval$ seconds. A pull at time
$t$ outputs a dataframe $L_{t}$ of tweet-IDs and their liking users.
Further, it continuously outputs dataframes $T_{t}$ that contain
tweets, like count, retweet count, and meta-data including time of
origin, text, posting user, language etc. Alg.~\ref{alg:Part-1-perpetualcollection}
and Alg.~\ref{alg:Part-2-=00005Bfinalharvest=00005D} require the
input parameters in Table~\ref{tab:Input-parameters}.
\begin{table}[H]
\begin{tabular}{|>{\raggedright}p{1\columnwidth}|}
\hline
\begin{lyxlist}{00.00.0000}
\item [{$keyword$}] Keyword(s) or hashtag(s). e.g. \texttt{\#dkpol}.\vspace{7pt}
\item [{$token,\;|token|$}] ARA Twitter Authentication Bearer Token, number
of tokens. More than 1 is possible. More raise request limits.\vspace{7pt}
\item [{$startpoint$}] Date and time to start data collection. Must be
in the past. E.g now, minus 10 seconds.\vspace{7pt}
\item [{$observationtime$}] Observation period. E.g., 1 hour, or 60 days.\vspace{7pt}
\item [{$tracktime$}] How long to track each tweet for new likes. E.g.,
48 hours,. Longer periods use up rate limit more quickly.\vspace{7pt}
\item [{$pullinterval$}] Sleep interval between pull completion and next
pull. E.g. 300 seconds. Shorter interval use up rate limit more quickly.\vspace{7pt}
\item [{$min.delta$}] How many new likes must a tweet have gotten before
we request its liking users? To play safe, satisfy $min.delta+min.delta\leq req.rate.lim$.\vspace{7pt}
\item [{$top.n$}] Determines from how many tweets to request likers per
pull. To play safe, satisfy $top-n\leq rlim\cdot\frac{pullinterval}{15\cdot60sec}\cdot|token|$.\vspace{7pt}
\item [{$min.likes$}] Minimum like (retweet) count of tweets to be considered
for final harvest. E.g. 1 or 10.\vspace{7pt}
\item [{$req.rate.lim$}] Twitter rate limit: 75 requests per 15 min. for
liking and retweeting users each.
\end{lyxlist}
\tabularnewline
\hline
\end{tabular}

\caption{\label{tab:Input-parameters}Input parameters for Algorithms \ref{alg:Part-1-perpetualcollection}
and \ref{alg:Part-2-=00005Bfinalharvest=00005D}.}
\end{table}
\begin{algorithm}
\begin{algor}[1]
\item [{{*}}] \textbf{Input: $keyword,token,startpoint,observationtime,$
}$pullinterval,tracktime,min.delta,top.n$
\item [{{*}}] \textbf{Output:} $T_{t},L_{t}$ for $t\in pullpoints:={\{}t\leq endpoint\colon\,\,\,t=startpoint+k\cdot pullinterval\text{ for a }k\in\mathbb{N}{\}}$
\item [{if}] exists file $log$
\item [{{*}}] load $log$ // to resume from error
\item [{else}]~
\item [{\texttt{{*}}}] $log\leftarrow\varnothing$ // start empty
dataframe with columns $tweet,like.count,like.count.last$ to track
tweets' like count now and last their likers were pulled
\item [{endif}]~
\item [{while}] true
\item [{if}] $sys.time=t$ for some $t\in pullpoints$ // if now is a time
to pull
\item [{{*}}] $T_{t},L_{t}\leftarrow\varnothing$ // start empty dataframes
for tweets and their metadata, and for liking users
\item [{{*}}] $start=\text{\ensuremath{\begin{cases}
startpoint\textbf{ if }t-tracktime<startpoint\\
t-tracktime\text{\textbf{ else} }
\end{cases}}}$
\item [{{*}}] $end=\text{\ensuremath{\begin{cases}
t\text{ \textbf{if} }t<startpoint+observationtime\\
startpoint+observationtime\textbf{ else}
\end{cases}}}$
\item [{{*}}] $T_{t}\leftarrow\mathsf{get\_tweets}(keyword,start,end,token)$
// pull tweets (incl. $like.count$) under $keyword$ posted between
$start$ and $end$, auth. with $token$
\item [{{*}}] save $T_{t}$ // save to file with timestamp
\item [{{*}}] $log\leftarrow\mathsf{update\_log\_1}(log,T_{t})$ // For
$tweet$ in $T_{t}$: if $tweet$ is not in $log$, append it with
$like.count$ from $T_{t}$ and $like.count.last=0$; else update
$tweet$'s $like.count$ in $log$ to its $like.count$ in $T_{t}$
\item [{{*}}] $candidates\leftarrow\mathsf{find\_candidates}(log,min.delta)$
// return list of all $tweet$ in $log$ for which $delta:=like.count-like.count.last\geq min.delta$
\item [{{*}}] sort $candidates$ by $delta$ in descending order // introduce
retrieval priority.
\item [{{*}}] $top\leftarrow candidates[0:top\_n-1]$ // restrict to $top\_n$
tweets with highest $delta$.
\item [{for}] $tweet$ in $top$
\item [{{*}}] $L_{t}\leftarrow\text{\ensuremath{\mathsf{get\_likers}}}(tweet,token)$
// pull 100 most recent likers
\item [{{*}}] $log\leftarrow\mathsf{update\_log\_2}(log,T_{t})$ // update
$tweet$'s $like.count.last$ in $log$ to its $like.count$ in $T_{t}$
\item [{endfor}]~
\item [{{*}}] save $L_{t}$ // save to file with timestamp
\item [{{*}}] save $log$ // save to file
\item [{else}]~
\item [{{*}}] break
\item [{endif}]~
\item [{endwhile}]~
\end{algor}
\caption{\label{alg:Part-1-perpetualcollection}Main loop of algorithm to retrieve
liking users from Twitter}
\end{algorithm}

\begin{algorithm}[t]
\begin{algor}[1]
\item [{{*}}] \textbf{Input:} $token,min.likers,req.rate.lim,\boldsymbol{T}={\{}T_{t}\colon\text{ output dataframe of Alg. \ref{alg:Part-1-perpetualcollection}}{\}}$
\item [{{*}}] \textbf{Output:} $L_{final}$
\item [{{*}}] $L_{final}\leftarrow\varnothing$ // Start empty dataframe
columns $tweet,last.likers$
\item [{{*}}] $all\leftarrow\mathsf{all\_tweets}(\boldsymbol{T})$ // Load
and concatenate all $T_{t}$. For duplicates, keep tweets with highest
$like.count$\emph{. }Subset columns of $all$ to $tweet$ and $like.count$,
rows to those with $like.count\geq min.likers$
\item [{{*}}] $counter\leftarrow0$
\item [{for}] $tweet$ in $all$
\item [{{*}}] $L_{final}\leftarrow\mathsf{get\_likers}(tweet,token)$ //
Pull 100 most recent likers, append to $L_{final}$
\item [{{*}}] $counter=counter+1$
\item [{if}] $counter\geq req.rate.limit\cdot|token|$
\item [{{*}}] $counter=0$
\item [{{*}}] sleep for $15$ minutes // Reset request limits
\item [{endif}]~
\item [{endfor}]~
\item [{{*}}] save $L_{final}$ // Save to external file
\end{algor}
\caption{\label{alg:Part-2-=00005Bfinalharvest=00005D}Final harvest to retrieve
liking users from Twitter}
\end{algorithm}

\section{Case Study: Data Collection and Analysis of the Danish Twittersphere\label{sec:Case-Study:-Data}}

To study both the performance of the contributed script and the usefulness
of the resulting dataset to address CIB, we analyze a case study of
the Danish political Twittersphere.

\subsection{Dataset: Parameters, Completeness and Descriptive Statistics}

The dataset used in this paper was collected using the described
script, without manual intervention during its runtime. The text query
was ``\#dkpol -is:retweet'', meaning that the script sought tweets
falling under \texttt{\#dkpol}, excluding retweets. Two bearer tokens
were used, doubling the request resources available. The observation
period started the afternoon of May 25th, 2022 and was 30 days long.
Tweets were tracked for 48 hours, as prior tests had shown that liking
activity on almost all tweets under \texttt{\#dkpol} stops before
48 hours after posting. The interval between pulls was 5 minutes,
and each pull requests the liking users of the top $top.n=36$ tweets
with $min.delta=3$.

Following the observation period, we requested the last 100 most recent
liking users of all tweets that had at least $10$ likes. We used
this limit to strongly diminish the amount of tweets in the final
check, with the justification that that so little total liking activity
would most likely not be hurtful coordinated inauthentic behavior.
In total, the script collected $47,714$ liking user IDs for $13,243$
tweets. While this case study focuses on liking behavior, the published
dataset contains retweeting user IDs as well.

\begin{figure}
\resizebox{1\columnwidth}{!}{\includegraphics{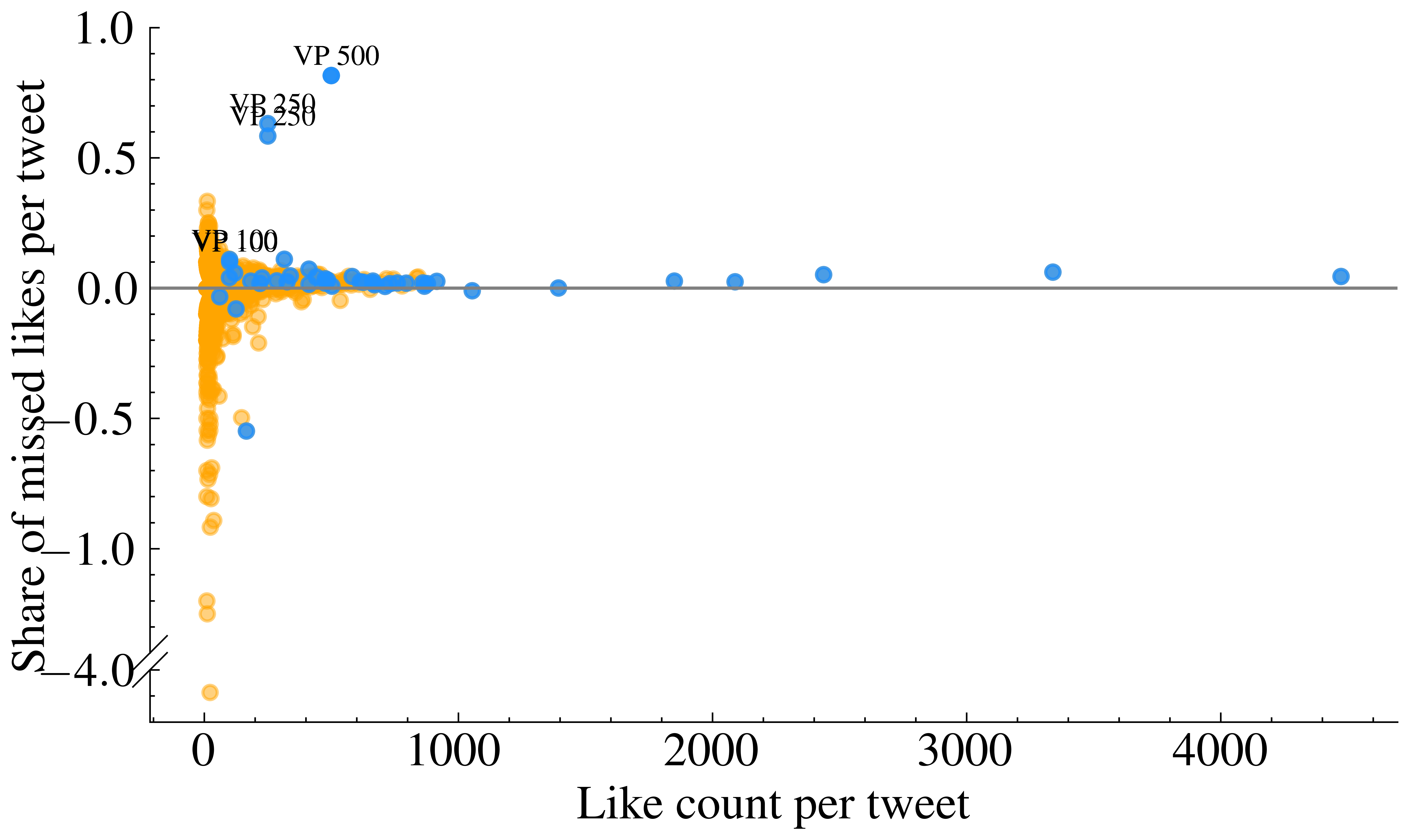}}\caption{\label{fig:completeness}Missed likes per tweet, as share of its maximal
like count, arranged by like count in ascending order. Dots represent
tweets. Labels ``VP $n$'' are on tweets for which we vendor-purchased
$n$ likes. Blue marks the tweets of the 50 largest bins of perfectly
correlated likes (cf. Sec~\ref{subsec:CaseStudy:}).}
\end{figure}

To assess completeness of the dataset, we compare the number of collected
likers (for those tweets subject to final harvest collection) to the
maximum like count a tweet has logged during the tracktime for each
tweet, i.e. 48 hours, see Fig~\ref{fig:completeness}.

First, we see both positive \emph{and negative }deviation numbers.
Positive deviation is expected: the script cannot collect more than
100 liking users per tweet per pull interval. For testing, we used
vendor-purchased likes on clearly-marked test-tweets. We highlight
the targeted tweets in Fig~\ref{fig:completeness}. As we purchased
some batches of more than $100$ likes and these were placed almost
simultaneously (some vendors place likes more slowly), we miss out
on collecting them. To detect CIB, this is not necessarily a problem:
temporal detection methods leveraging time-synchronous user behavior
to detect coordination can easily identify such behavior. Negative
deviation indicates that the script has collected more liking users
than the \emph{like} \emph{count} suggests. This happens when likes
are retracted, the liking profiles are deleted,\footnote{These are both actions genuine users may take, but are also often
observed with vendor-purchased metric inflation \cite{Torres-Lugo_Likes_Manip_deletions}.} or a tweet attracted likes post tracktime, which we collected in
the final harvest.

Second, we find that for high engagement tweets, the script performs
well and collects most of the liking users. In contrast, for very
low engagement tweets, the script is more prone to miss out on more
than $10\%$ of users. This is due to the algorithm prioritizing tweets
that get traction by allocating requests to collect the growing sets
of likers.

Third, and to complement the plots in Fig.~\ref{fig:completeness},
for $39.98\%$ of $6702$ tweets, the script collects exactly as many
liking users as the like count suggests. For $93.7\%$ of the $6702$
tweets, the script collects numbers of likers that fall within $10
$ of the like count. If considering negative deviation only, in $96.6\%$
of $6702$ tweets, the script deviates negatively $10\%$ or less.
If considering positive deviation only, in $97.06\%$ of $6702$ tweets,
the script deviates positively $10\%$ or less. I.e., in $97\%$ of
cases, the script seemingly collects $90\%+$ of liking users.

\subsection{Analysis: Perfect Correlation\label{subsec:CaseStudy:}}

In this case study, we make use of very simple user data: a binary
matrix containing a row for each tweet and column for each user, each
cell marked $1$ if the user liked the tweet, else $0$. Again, the
dataset contains temporal data as well, but we ignore it here, as
we are mainly interested in seeking patterns in like behavior alone.

Assume we have observed $n$ tweets. Let $Likers_{k}$, $k\leq n$,
be the set of users observed to have liked tweet $k$, so $Likers=\cup_{k\leq n}Likers_{k}$
is the set of all observed liking users. With $m=|Likers|$, we then
compress our data to a binary $n\times m$ matrix with entry values
in $\{0,1\}$, each row representing a tweet, each column a user.
With this matrix called $\mathbf{L}$, the entry $\mathbf{L}_{i,j}=1$
if user $i$ has liked tweet $j$, and $0$ else. Henceforth, we hence
identify user $i$ with the row $\mathbf{L}_{*,i}$ that contains
their like profile. In this case study, $\mathbf{L}$ is of dimension
$13,243\times47,714$.

We seek to group users as exhibiting coordinated liking behavior if
their like profiles are sufficiently similar, according to some measure.
Existing work routinely projects bipartite data structures (which
$\mathbf{L}$ is) onto a user-user similarity graph using a distance
or similarity metric (e.g., \cite{Pacheco2021Coordinated,nizzoli2021coordinated})
or develops algorithms to detect dense subgraphs to identify anomalous
groups of nodes (e.g., \cite{hooi2020telltail,shin2017densealert}).
Here, we apply the strictest measure: we group two users if, and only
if, they exhibit \emph{exactly the same like behavior}. This is equivalent
to grouping users that have cosine similarity $1$, Jaccard similarity
$1$, or Hamming distance $0$.

We apply this strictest measure as behavior labeled as coordinated
will also be labeled as coordinated using any less discriminating
measure. The approach thus is precautious with regard to labeling
coordinated users. The method is not designed to identifying all coordinated
inauthentic behavior in likes. There may very well be nuances and
less than perfectly correlated inauthentic behavior. To answer whether
a collection of tweet likes exhibits first signs of CIB, we propose
the method only as valid for positive answers: if this strongly discriminatory
methods finds such signs, then methods with lower bars for coordination
should, too. If the method does not find such signs, we would deem
it fallacious to take this as evidence that no CIB occurred.

To group users with identical like profiles, we worst case have to
pair-wise compare all users, i.e. undertake $\frac{47,713^{2}-47,714}{2}$
comparisons. To avoid as many of these comparisons as possible, we
sort users into bins: we initiate a list with one bin containing the
first user. For every later user, we compare them with one user from
each bin in the list of bins, checking larger bins first, and stopping
to place them in the first bin that provides a perfect match. If no
such bin exists, we add a new bin for the user in the end of the list.
We find only $25,806$ bins.

$49.9\%$ of users are sorted into bins of size 1. Filtering for
bins of at least size 50 (as smaller bins are negligible in impact
for CIB), we find $50$ bins with $13,018$ out of $47,714$ users.
Put differently, $27.28\%$ of users are in a group with at least
$49$ others that share the exact same like behavior across all $13,243$
tweets. These $27.28\%$ like most often only $1$ tweet, sometimes
$2$. In the largest bin, $3,217$ users are perfectly correlated
liking the same tweet.

\begin{figure}
\begin{centering}
\resizebox{1\columnwidth}{!}{\includegraphics{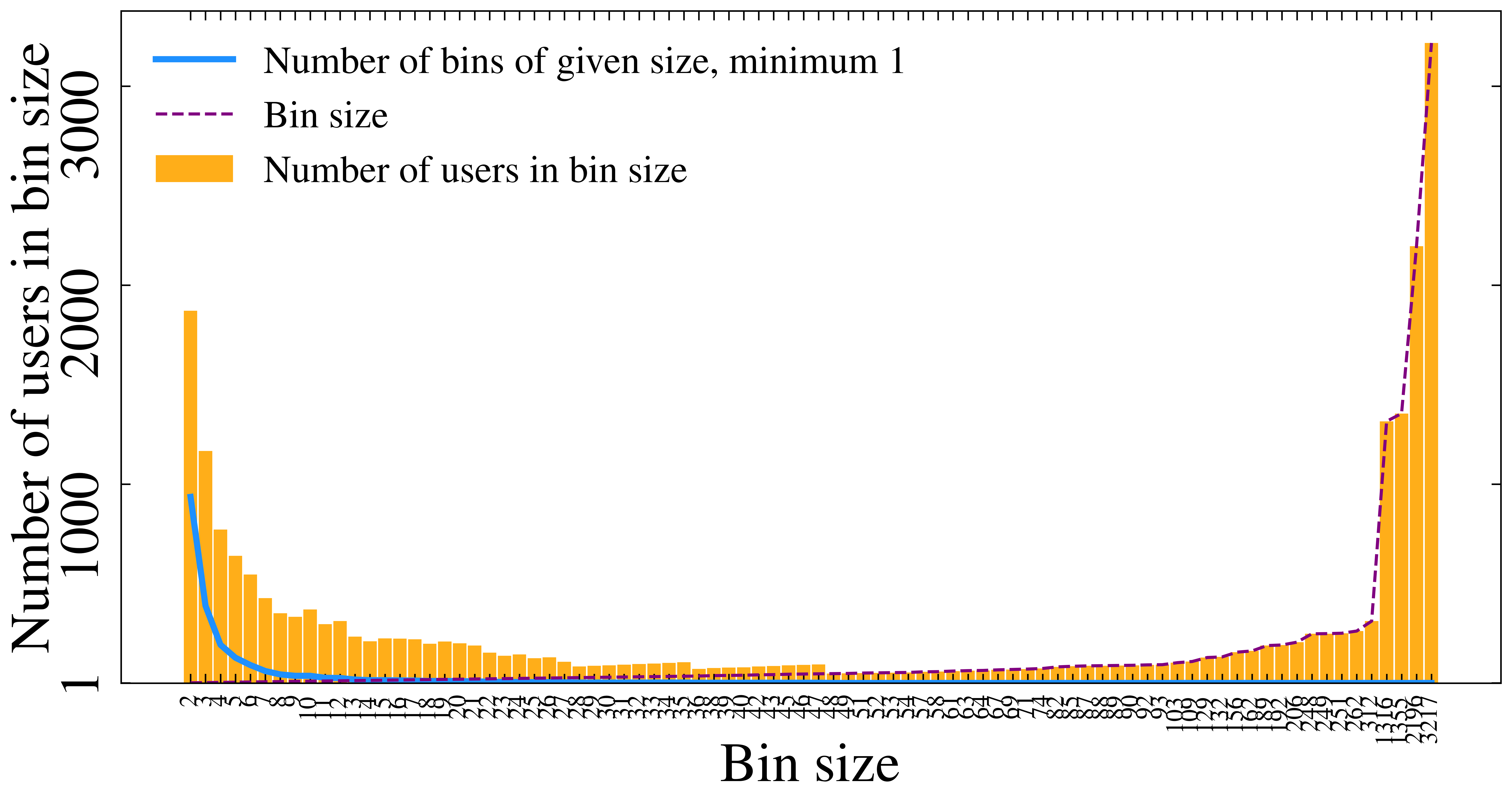}}
\par\end{centering}
\caption{\label{fig:Bins} Bins with at least two users, the number of users
in bins of each size, and the number of bins of each size. E.g., the
left-most bar shows there are $\sim2000$ users (yellow bar) distributed
over $\sim1000$ bins (solid blue line) of size $2$ (dotted purple
line), while the right-most shows there are $\sim3217$ users distributed
over $1$ bin of size $3217$. The number of bins of size $n$ drops
to 1 at $n=48$.}
\end{figure}

Collected in their own bins, we find the users behind the likes we
purchased from online vendors. We refer to Fig.~\ref{fig:Bins} for
an overview of the magnitude of bins.

We find several bins of users with perfectly identical liking behaviors
unrelated to our purchases. We cannot conclude from \emph{correlation}
to \emph{coordination} to state these bins contain users engaged in
coordinated inauthentic behavior. We do find the larger bins suspicious
and in warrant of further analysis, cf. the discussion in Sec.~\ref{sec:Discussion-and-Ethical}.

We find the larger bins suspicious as we find it unlikely that the
correlation has arisen without coordination. E.g., rate the probability
of each bin as being non-coordinated using the following charitable
assumptions (charitable to favor the odds of large bins): Assume
that the probability that any two users share the exact same like
profile without being coordinated is $c=.95$. For simplicity and
charity, ignore that this probability attaches to every unordered
pair of users in a bin, and let the probability that a bin $B$ of
size $|B|$ occurred without coordination be $P(B)=c^{|B|-1}$, i.e.,
the probability that $|B|-1$ users pairwise and independently correlated
with the same user $i$ from $B$. This probability drops drastically
with the growth of $B$:\medskip{}

\begin{center}
\begin{tabular}{c|c|c|c|c|c|c|c}
$|B|=$ & 2 & 10 & 50 & 60 & 75 & 100 & 200\tabularnewline
\hline
$P(B)=$ & .95 & .63 & .08 & .05 & .02 & .006 & $3.69\cdot10^{-5}$\tabularnewline
\end{tabular}\medskip{}
\par\end{center}

\noindent These (fictitious) probabilities do not mean that it is
unlikely that e.g. $60$ users liked the same tweet\textemdash but
that it is unlikely that they all liked or did not like \emph{all
the same tweets}. Even under charitable conditions, bins larger than
$60$ quickly seem highly unlikely. We further discuss the implications
of our results in Sec.~\ref{sec:Discussion-and-Ethical}.

\paragraph{Singular Value Decomposition}

To visualize and locate the identified bins among all users, we turn
to plotting the data in a dimensionality-reduced space: With dimensionality
reduction, user behavior often exhibits a clustered structure, for
example, separating bots and humans in labeled bot datasets \cite{Yang2020,Nwala2022BLOC},
disclosing synchronous clusters of retweeters \cite{giatsoglou2015nd}
(later used in baseline experiments by \cite{dutta2018retweet,chetan2019corerank,arora2020analyzing}),
revealing generally correlated groups such as polarized groups of
users \cite{wojcik2022birdwatch_twitter} among users writing Twitter
Birdwatch notes, or coordinated clusters of agents as in \cite{MCSpaper}
given computer-simulated data.

\begin{figure}
\begin{centering}
\resizebox{1\columnwidth}{!}{\includegraphics{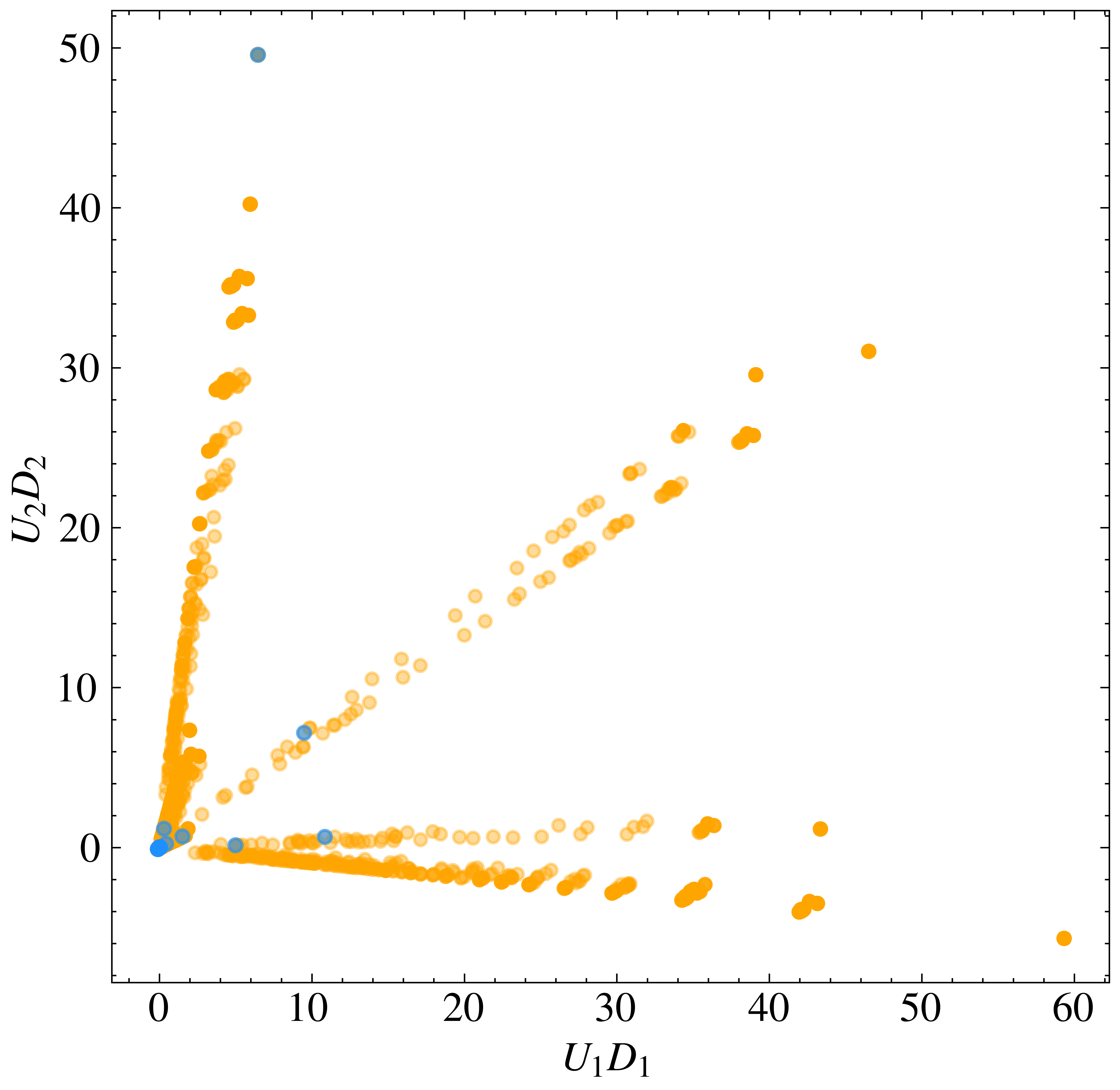}}
\par\end{centering}
\caption{\label{fig:Missedlikes-1-1} Scatterplot of $\mathbf{U}_{q}\mathbf{D}_{q}$.
Top 50 perfectly correlated bins of users overlap perfectly with one
another in clusters colored in blue. Bins of vendor-purchased likers
are among the bottom left groups of clusters.}
\end{figure}

We calculate the singular value decomposition (SVD)\linebreak{}
 $\mathbf{X=UDV}^{T}$ of the $m\times m$ sample correlation matrix
$\mathbf{X}$ of the data in matrix $\mathbf{L}$. We consider the
first $q=2$ dimensions' eigenvectors, i.e., the first two columns
of the $n\times p$ orthogonal matrix $\mathbf{U}$ where $n=p$,
weighted with the corresponding eigenvalue collected in the diagonal
$p\times p$ matrix $\mathbf{D}$ \cite{hastie_09_elements-of.statistical-learning}.
We plot the scatterplot of $\mathbf{U}_{q}\mathbf{D}_{q}$ in Figure~\ref{fig:Missedlikes-1-1}.
 In the plot, each dot represents a liking user. While we color-coded
the users placed in the largest 50 bins, they may also be discerned
through their darker shade that stems from many dots perfectly overlapping
one another. The SVD and the scatterplot thus picks up on correlation
and the vendor-purchased metric inflation. As an alternative route,
note that clustering on these first two eigenvectors (e.g. using a
Gaussian Mixture Model as done in \cite{MCSpaper}) picks up on the
inauthentically coordinated users we know of, too.

\section{Concluding Remarks: Discussion \& Ethical Considerations\label{sec:Discussion-and-Ethical}}

\paragraph{Data collection discussion}

The script we have presented here is designed to collect the IDs of
liking (and/or retweeting) users of tweets that satisfy a selected
textual query. As such, the script takes a \emph{domain first }perspective
on data collection, rather than a \emph{user first }perspective as
most other work designed to investigate coordinated inauthentic behavior.

The dataset presented in this paper is collected around the domain
of the Danish political Twittersphere, found under \texttt{\#dkpol}.
For this domain, using the parameters described and two bearer tokens,
the script had a reasonably low rate of missing liking users, and
misses more than $10\%$ of liking users in only $3\%$ of cases when
run continuously for 30 days. Such a targeted dataset cannot be obtained
directly through any of Twitter's data access options.

In an international context, \texttt{\#dkpol} is a small domain. With
the same parameters and number of bearer tokens, the script would
indubitably fare less well on much larger domains. For larger domains
with more intense liking activity, it would be interesting to study
the script's performance with more bearer tokens and far more aggressive
pull parameters, such as much lower $pullinterval$. As data retrieval
from Twitter is not instantaneous (especially when it comes to updating
the like count of a large batch of tweets), we suspect that a satisfactory
data collection will involve multiple machines running the script
in parallel, each tracking a subset of tweets assigned to them (e.g.
using tweet ID \emph{modulo $k$ }for $k$ machines).

Another, and favorable, option for obtaining the data on one-click
reactions would be if Twitter or other social media platforms made
this data available to the research community. We hope that the case
study in this paper\textemdash where even a crude and strict analysis
raises red flags for CIB\textemdash may be used as an argument that
one-click reaction data is relevant in the study of coordinated inauthentic
behavior and thus in the arms race against online misinformation to
ultimately put pressure on the social media industry to release data.

\paragraph{Analysis discussion}

In our case study, the controlled CIB through vendor-purchased likes
is grouped into distinct bins that we can match to our tweets. The
coordination here is achieved through weak ties in our bipartite graph
structure $\mathbf{L}$. We complement, for example, Weber et al.'s
\cite{weber2021amplifyingcoord} approach focused on coordination
through strong ties. As \cite{weber2021amplifyingcoord} acknowledges
as an open issue and we show, coordination may take place along weak
ties. With our like-based approach, we provide first steps towards
a measure to detect such. In contrast to existing work (e.g. \cite{Schoch2022}),
the present like-based approach does\emph{ not} need to filter the
data for strongly tied communities, highly influential users and superspreaders,
or very active or users that, e.g. like a minimum number of times
within a short period. Without filtering, we are able to group users
with such behaviors together.

Our analysis made use of vendor-purchased likes. Purchasing engagement
metric inflations violates Twitter's platform manipulation and spam
policy \cite{TwitterPlatformManip}, which defines ``platform manipulation
as using Twitter to engage in bulk, aggressive, or deceptive activity
that misleads others and/or disrupts their experience.''. We created
two Twitter accounts that in the name of the research center with
which the authors are affiliated (`CIBS1' (@CIBS110) and `CIBS2'
(@CIBS22)) posted $6$ tweets with text `\emph{Research test tweet
$n$/6. Apologies for spamming \#dkpol.}' for $n=1,...,6$. We inflated
the like count for these $6$ tweets. We acknowledge that the coordinated
inflation of these tweets might have disrupted the experience of Twitter
users. To the best of our assessment, the amplification of these tweets
does not comprise \emph{harmful} coordinated activity nor was it deceptive
or commercially-motivated, but declared a research motivation. Ethically,
we thus believe that the benefits of studying coordinated inauthentic
behavior outweigh the minimal disruptions we have caused to Twitter
users by violating Twitter's manipulation and spam policy.

Unrelated to our purchases, we further find and visualize several
large groups of users with perfectly correlated, identical liking
behaviors\textemdash similarly achieved through weak ties. We have
no ground truth about whether the suspected accounts beyond our test
are naturally correlated and not inauthentically coordinated, yet
we believe that natural correlation is unlikely enough that such groupings
are red flags for CIB, and warrant further inspection, out of scope
of this case study. Our methods may thus serve as pre-studies for
bot detection and the application of fact checkers \cite{Kirn2022}.
Further, the dataset and explorative case study may serve as a point
of departure for future research to explore the correlation structures
among liking users and the development of novel detection methods.

\paragraph{Censorship}

Any flagging of behavior in public fora raises ethical concerns about
censorship. The classification of reactions such as likes and retweets
to tweets is no different. Generally, we find that the flagging of
coordinated behavior used by inauthentic attention hackers is defendable,
justified by the aim to combat misinformation online. We omit further
discussion of this point. However, in applying automated techniques,
there is always a risk of misclassification. If a technique is used
for censorship, this may lead to unrightful labeling. The methods
for initial exploration proposed here may then risk unjustified labeling
users due to behavioral correlation with strongly coordinated groups
of users. We strongly recommend that the methods here are taken as
a first step towards fact-checking content and users and not as a
final verdict about specific individual users.

\paragraph{Data collection approval}

Approval of data collection and processing of personal data in the
research project was granted by the faculty secretariat of the university
of Copenhagen. The approval emphasizes that the processing of personal
data in the project is in accordance with the rules of the European
General Data Protection Regulation, Regulation 2016/679 on the protection
of natural persons with regard to the processing of personal data.
That the study would be undertaken was made public on the authors'
university websites.

\paragraph{Datasets and code availability}

Dataset and code are made available for the research community \cite{Dataverse},
hosted on the archival repository Harvard Dataverse that provides
a Document Object Identifier (DOI) for better findability. To comply
with the Twitter terms, access to the data on Harvard Dataverse is
granted when researchers actively agree to the Twitter Terms of Service,
Privacy Policy and Developer Policy. The data collection code is also
available on the public GitHub repository\emph{ Get-Twtter-Likers-Data}
\cite{JahnRendsvig22Git_likers}.

\end{document}